\documentclass[12pt]{article}
\usepackage{graphicx,psfrag,epsf,color}

\newcommand{\bff}[1]{\mbox{\boldmath ${#1}$}}

\catcode`@=11
\newcount\@tempcntc
\def\@citex[#1]#2{\if@filesw\immediate\write\@auxout{\string\citation{#2}}\fi
  \@tempcnta\z@\@tempcntb\m@ne\def\@citea{}\@cite{\@for\@citeb:=#2\do
    {\@ifundefined
       {b@\@citeb}{\@citeo\@tempcntb\m@ne\@citea\def\@citea{,}{\bf ?}\@warning
       {Citation `\@citeb' on page \thepage \space undefined}}%
    {\setbox\z@\hbox{\global\@tempcntc0\csname b@\@citeb\endcsname\relax}%
     \ifnum\@tempcntc=\z@ \@citeo\@tempcntb\m@ne
       \@citea\def\@citea{,}\hbox{\csname b@\@citeb\endcsname}%
     \else
      \advance\@tempcntb\@ne
      \ifnum\@tempcntb=\@tempcntc
      \else\advance\@tempcntb\m@ne\@citeo
      \@tempcnta\@tempcntc\@tempcntb\@tempcntc\fi\fi}}\@citeo}{#1}}
\def\@citeo{\ifnum\@tempcnta>\@tempcntb\else\@citea\def\@citea{,}%
  \ifnum\@tempcnta=\@tempcntb\the\@tempcnta\else
   {\advance\@tempcnta\@ne\ifnum\@tempcnta=\@tempcntb \else \def\@citea{--}\fi
    \advance\@tempcnta\m@ne\the\@tempcnta\@citea\the\@tempcntb}\fi\fi}
\catcode`@=12

\begin{document}

\begin{titlepage}

\begin{flushright}
PITHA 07/02\\
TTP/07-10\\
SFB/CPP-07-21 \\
0705.4518 [hep-ph]\\
29 May 2007
\end{flushright}

\vskip1.5cm
\begin{center}
\Large\bf\boldmath
Third-order non-Coulomb correction to the S-wave quarkonium wave
functions at the origin
\end{center}

\vspace{1cm}
\begin{center}
{\sc M.~Beneke}$^a$,
{\sc Y. Kiyo}$^b$,
{\sc K. Schuller}$^a$\\[5mm]
  {\it $^a$ Institut f{\"u}r Theoretische Physik~E,
    RWTH Aachen,}\\
  {\it D--52056 Aachen, Germany}\\[0.1cm]
  {\it $^b$ Institut f{\"u}r Theoretische Teilchenphysik,
    Universit{\"a}t Karlsruhe,}\\
  {\it D--76128 Karlsruhe, Germany}\\[0.3cm]
\end{center}

\vspace{2cm}
\begin{abstract}
\noindent
We compute the third-order correction to the $S$-wave
quarkonium wave functions $|\psi_n(0)|^2$ at the origin from
non-Coulomb potentials in the effective non-relativistic
Lagrangian. Together with previous results on the Coulomb
correction and the ultrasoft correction computed in
a companion paper, this completes the third-order calculation
up to a few unknown matching coefficients. Numerical
estimates of the new correction for bottomonium and
toponium are given.
\end{abstract}
\end{titlepage}

\section{Introduction}

The non-relativistic bound-state problem has a long history
since the birth of quantum mechanics. Its systematic derivation from
the relativistic quantum field theory of electrodynamics or
chromodynamics (QCD) was developed more recently. 
Non-relativistic effective field
theories~\cite{Caswell:1985ui,Pineda:1997bj,Luke:1999kz} together
with dimensional regularization and diagrammatic expansion
methods~\cite{Beneke:1997zp} now allow calculations of higher-order
perturbative corrections, including all ``relativistic'' effects,
to the leading-order bound-state properties, given by
the solution of the Schr\"odinger equation. This is of interest
in QCD for the lowest bottomonium state and top-antitop production
near threshold, where non-perturbative long-distance
effects can be argued to be sub-dominant, but perturbative
corrections are large.

The $S$-wave energy levels are currently known at
next-to-next-to-next-to-leading order 
(NNNLO)\footnote{Non-relativistic perturbation theory is an 
expansion in $\alpha_s$ and the non-relativistic velocity $v$, while 
counting $\alpha_s/v \sim 1$, which implies a summation of the series 
in $\alpha_s$ even at LO. We do not sum logarithms of 
$\alpha_s \ln v$.}~\cite{Kniehl:1999ud,Kniehl:2002br,Beneke:2005hg,Penin:2005eu}, except
for the three-loop coefficient of the colour-Coulomb potential, but
the corresponding wave functions at the origin, which are related to
electromagnetic decay and 
production of these states are completely known only
at next-to-next-to-leading order
(NNLO)~\cite{Melnikov:1998ug,Penin:1998kx,Beneke:1999qg}.
There exist partial results for logarithmic effects at
NNNLO~\cite{Kniehl:1999mx,Manohar:2000kr,Kniehl:2002yv,Hoang:2003ns}, 
which can be
related to certain anomalous dimensions and lower-order quantities.
In \cite{Beneke:2005hg} we computed the third-order
corrections to $S$-wave wave function at the origin from
all terms in the heavy-quark potential related only to the
Coulomb potential. In this paper we compute the contribution
from the remaining potentials. A companion paper \cite{Beneke:2007bkp}
deals with the Lamb-shift like contribution from ultrasoft
gluons, thus completing the calculation of all bound-state
effects at NNNLO, except for a few unknown matching coefficients. Our
result is provided in such a form that these coefficients can be
easily inserted, once they are computed. 

In contrast to the Coulomb corrections the calculation of
the more singular non-Coulomb potential corrections leads to
divergences, both in the calculation of the potentials themselves
as in the insertions of these potentials in the calculation
of the wave function at the origin. We employ dimensional
regularization with $d=4-2\epsilon$
throughout, and provide
a precise definition of all quantities, which corresponds to
the $\overline{\rm MS}$ subtraction scheme. 
The technical details of this calculation together with an
extension to the full $S$-wave Green function will be given
elsewhere.

\section{Relating the leptonic quarkonium decay constant
to the wave function at the origin}

We consider the two-point function
\begin{eqnarray}
\label{eq:TwoPointFunction}
&&
\left(q^\mu q^\nu - g^{\mu\nu} q^2 \right)\Pi(q^2)
= i \int d^d x\, e^{i q x}\,
\langle \Omega | T(j^\mu(x)j^\nu(0)) |\Omega\rangle,
\end{eqnarray}
of the electromagnetic heavy-quark current $j^\mu=\bar{Q}\gamma^\mu
Q$, choosing $q^\mu=(2 m+E,\bff{0})$ with $m$ the
pole mass of the heavy quark. The two-point function exhibits
the $S$-wave bound-state poles at $E_n$, near which
\begin{equation}
\Pi(q^2) \stackrel{E\rightarrow E_n}{=}
\frac{N_c}{2 m^2}\,\frac{Z_n}{E_n-E-i \epsilon}.
\end{equation}
Here $N_c=3$ denotes the number of colours.
The residue $Z_n$ is related to the leptonic decay width
$\Gamma([Q\bar{Q}]_n\to l^+ l^-)$
of the $n$th $S$-wave quarkonium state by
\begin{eqnarray}
\Gamma([Q\bar{Q}]_n\to l^+ l^-)
&=& \frac{4\pi N_c e_Q^2\alpha^2 Z_n}{3 m^2},
\end{eqnarray}
with $e_Q$ the electric charge of the heavy quark in units of the
positron charge, and $\alpha$ the fine-structure constant. Although
there are no toponium states, and the cross section of
top-antitop production is determined by the full
two-point function, the residue $Z_n$ for $n=1$ provides
an approximation to the height of the broad resonance in
this cross section.

The electromagnetic current $j^\mu$ is expressed in terms of 
the non-relativistic
heavy quark ($\psi$) and anti-quark ($\chi$) field operators via
\begin{eqnarray}
\label{eq:QCDVectorCurrent}
j^{\,i}=c_v\, \psi^\dag\sigma^i\chi
+ \frac{d_v}{6m^2}\psi^\dag\sigma^i\,{\bf D^2}\chi
+\ldots,
\end{eqnarray}
where the hard matching coefficients have expansions
$c_v=1+ \sum_{n} c_v^{\,(n)} (\alpha_s/4\pi)^n$,
and the $d_v=1+ d_v^{\,(1)} (\alpha_s/4\pi)+\ldots$.
The central quantity in this paper is the two-point function
\begin{eqnarray}
\label{eq:G}
G(E)&=&\frac{i}{2 N_c (d-1)} \int d^{d} x\, e^{iEx^0}\,
\langle\Omega| T(\,
[\psi^{\dag}\sigma^i\chi](x)\,
[\chi^{\dag}\sigma^i\psi](0))
|\Omega\rangle
\nonumber\\
&\stackrel{E\rightarrow E_n}{=}&
\frac{|\psi_n(0)|^2}{E_n-E-i \epsilon},
\end{eqnarray}
defined in non-relativistic QCD (NRQCD), whose poles define the
wave functions at the origin and bound-state energy levels. At leading
order, the wave functions and binding energies are given by 
$|\psi^{(0)}_n(0)|^2=(m C_F \alpha_s)^3/(8\pi n^3)$
and $E^{(0)}_n=-m(\alpha_s C_F)^2/(4 n^2)$, respectively
(here and below $C_F=(N_c^2-1)/(2N_c)=4/3, C_A=N_c=3$).
They receive perturbative corrections from higher-order heavy-quark 
potentials and dynamical gluon effects, hence $E_n=E_n^{(0)}\,(1+ \sum_{k} 
(\alpha_s/4\pi)^k e_k)$ and $|\psi_n(0)|^2 =
|\psi_n^{(0)}(0)|^2\,(1+ \sum_{k}(\alpha_s/4\pi)^k f_k )$.
Using an equation-of-motion relation,
we can replace ${\bf D^2}$ in (\ref{eq:QCDVectorCurrent}) by
$-m E$, and we obtain
\begin{eqnarray}
\label{eq:Z}
Z_n&=& c_v \left[ c_v - \frac{E_n}{m}\left(1+\frac{d_v}{3}\right)+
\cdots\right] |\psi_n(0)|^2,
\end{eqnarray}
where terms beyond NNNLO are neglected. Inserting the 
perturbative expansions and defining  
$Z_n=|\psi_n^{(0)}(0)|^2\,(1+ \sum_{k} (\alpha_s/4\pi)^k z_k)$, 
results in 
\begin{eqnarray}
z_1 &=& 2 c_v^{(1)}+f_1,
\\
z_2 &=& 2 c_v^{(2)}+ {c_v^{(1)}}^2 +2 c_v^{(1)} f_1+ f_2
-\frac{4}{3}\,\frac{16\pi^2 E_n^{(0)}}{m\alpha_s^2},
\label{z2}
\\
z_3 &=& 2 c_v^{(3)}+2  c_v^{(1)} \Big(c_v^{(2)}+f_2\Big) +
\Big(2 c_v^{(2)}+ {c_v^{(1)}}^2\,\Big)\,f_1 
+ f_3
\nonumber\\
&& -\,\frac{16\pi^2 E_n^{(0)}}{m\alpha_s^2}\left[\frac{d_v^{(1)}}{3}+
\frac{4}{3}\,(c_v^{(1)}+e_1+f_1)\right].
\label{z3}
\end{eqnarray}
Note that $e_k$, $f_k$ and $z_k$ depend on the principal
quantum number $n$ of the energy level, but we omitted
a corresponding index to keep the notation short.
The short-distance coefficients $c_v^{(1)}$, $c_v^{(2)}$
in the $\overline{\rm MS}$ scheme\footnote{The $\overline{\rm MS}$ 
scheme is defined by the loop integration measure 
$\tilde\mu^{2\epsilon} d^dk/(2\pi)^d$ with 
$\tilde\mu^2=\mu^2 \,e^{\gamma_E}/(4\pi)$ and subtraction of 
the pole parts in $\epsilon$.} are given 
in \cite{Beneke:1997jm,Czarnecki:1997vz}.
The third-order coefficient $c_v^{(3)}$ is not yet known 
except for the $n_f$ terms ($n_f$ is the number of quarks
whose mass is set to zero)~\cite{Marquard:2006qi}.
The one-loop correction to the
derivative current can be obtained by applying a spin-triplet
projection to the results given in \cite{Luke:1997ys},
which gives
\begin{equation}
d_v^{(1)} = -C_F\left[32 \ln\frac{\mu}{m}+\frac{16}{3}\right].
\end{equation}
Here the infrared-divergent part $d_v^{(1,\rm div)} = -16C_F/\epsilon$
is subtracted and added back to the ultrasoft calculation,
where it cancels an ultraviolet divergence~\cite{Beneke:2007bkp}.
The first- and second-order corrections $e_1$, $f_1$ and $f_2$
are summarized in \cite{Beneke:2005hg}.\footnote{In Eq.~(30) of
the published version of \cite{Beneke:2005hg}, 
the term $-E_n^{(0)}/m$ in (\ref{z2}) 
that comes from the expansion of $q^2$ around $4 m^2$ 
was incorrectly included in the formula for $f_2^{nC}$. The 
term $-13/(8 n^2)$ in Eq.~(30) should therefore 
read $-15/(8 n^2)$. This is corrected in [hep-ph/0501289v2].}
The key quantity in the present work is the third-order correction
$f_3 = f_3^C+f_3^{nC}+f_3^{\rm us}$, which we split into three
parts: $f_3^C$ from the Coulomb potential given in
\cite{Beneke:2005hg} (see also \cite{Penin:2005eu});
$f_3^{nC}$ accounting for all remaining potential insertions
calculated below; and the ultrasoft correction
$f_3^{\rm us} = 64\pi^2\delta^{us}\psi_n$ calculated
in~\cite{Beneke:2007bkp}. Note that $e_1$, $f_1$, $c_v^{(1)}$
are finite, but all other expansion coefficients
$c_v^{(2)}$, $c_v^{(3)}$, $d_v^{(1)}$, $f_2$ and
$f_3^C$, $f_3^{nC}$, $f_3^{\rm us}$ depend on a convention
for regularizing (and subtracting) various ultraviolet and
infrared divergences that arise in separating the contributions
from the different scales. These divergences cancel in the
expansion coefficients $z_k$, since $Z_n$ is an observable.
We regulate all ultraviolet (UV) as well as infrared (IR) divergences 
dimensionally, and adopt the $\overline{\rm MS}$ subtraction scheme, 
but we also check explicitly that the pole parts of
the various terms cancel in the sums (\ref{z3}) wherever possible.

\section{Potentials}
Starting from full QCD, potentials arise after integrating out 
hard and soft modes, and potential gluons \cite{Beneke:1998jj}.
The potential NRQCD (PNRQCD)~\cite{Pineda:1997bj} effective 
Lagrangian contains a series of instantaneous interactions 
(potentials) with matching coefficients computed to an appropriate 
order in perturbation theory, 
and the interactions of ultrasoft gluons. In this paper we 
do not consider the ultrasoft gluon interactions (see \cite{Beneke:2007bkp}), 
so the required Lagrangian is
\begin{eqnarray}
{\cal L} &=& \psi^\dag \Big(i\partial_0+\frac{{\bff{\partial^2}}}{2m}+ 
\frac{ {\bff\partial^4}}{8m^3}\,\Big)\psi
+\chi^\dag \Big(i\partial_0-\frac{{\bff
\partial^2}}{2m}-\frac{ {\bff\partial^4}}{8m^3}\Big)\chi
\nonumber \\
&+& \int d^{d-1} {\bf r} \, \Big[ \psi^\dag \psi \Big](x+{\bf r}) \,
\Big(V_0 (r)+ \delta V (r,{\bff \partial})\Big)\, \Big[\chi^\dag \chi\Big](x).
\label{pnrqcd}
\end{eqnarray}
In the first line relativistic corrections to the kinetic energy
are included; the second line shows the potentials with $\delta V$
the correction to the leading-order Coulomb potential $V_0$. The
general form of the potential is colour- and spin-dependent. 
In writing (\ref{pnrqcd}) we have made simplifications 
that apply to the computation of the Green function 
of colour-singlet, spin-triplet currents. We have also used that 
annihilation diagrams contribute to this Green function only from 
the fourth order. Since we work in dimensional regularization, 
the spin-triplet projection has to be carried out consistently 
in $d$ dimensions, which implies ${\rm tr} (\sigma^i\sigma^i) = 2
(d-1)$ for the Pauli matrices. Furthermore one must not use the 
$\epsilon$-symbol, and treat the commutator $[\sigma^i, 
\sigma^j]$ as irreducible. For the calculation the momentum space
representation of $\delta V$ is more convenient. It can be 
written in the form (${\bf q}={\bf p}-{\bf p^\prime}$) 
\begin{eqnarray}
\label{eq:potential}
\delta \widetilde{V}
&=&
-\frac{4\pi \alpha_{s}C_F}{\bf{q}^2}
\bigg[\,{\cal V}_C
       -{\cal V}_{1/m}\,\frac{\pi^2\,|\bf{q}|} {m}
       +{\cal V}_{1/m^2}\,\frac{\bf{q}^2}{m^2}
       +{\cal V}_{p}\,\frac{\bf{p}^2+\bf{p}^{\prime \,2}}{2m^2}\,
\bigg].
\end{eqnarray}
The strong coupling constant $\alpha_s=\alpha_s(\mu)$
is renormalized at the scale $\mu$ unless stated otherwise. 
The coefficients ${\cal V}_i$ can be 
expanded in $\alpha_s$,
\begin{eqnarray}
{\cal V}_{i}={\cal V}_{i}^{(0)}+\frac{\alpha_s}{4\pi}{\cal
V}_{i}^{(1)}+\bigg(\frac{\alpha_s}{4\pi}\bigg)^2{\cal
V}_{i}^{(2)}+O(\alpha_s^3),
\end{eqnarray}
with the superscript denoting the number of loops.
The tree-level Coulomb potential ${\cal V}_C^{\,(0)}$
is a leading effect and is therefore included in $V_0(r)$ in the
Lagrangian. For the calculation of the third-order non-Coulomb 
correction we need ${\cal V}_{1/m}$ to two loops, and 
${\cal V}_{1/m^2}$, ${\cal V}_{p}$ to one loop. We also 
need the one-loop Coulomb potential to calculate the 
double insertion of ${\cal V}_C^{(1)}$ with 
${\cal V}_{1/m^2}^{(0)}$ and ${\cal V}_{p}^{(0)}$. 

The potentials must be determined in $d$ dimensions, since 
the subsequent insertions are ultraviolet divergent. 
We now summarize the expansion coefficients. 
The tree-level coefficients are \cite{Beneke:1999qg}
\begin{eqnarray}
{\cal V}_{p}^{\,(0)}=1,
\hspace{.4cm}
{\cal V}_{1/m}^{\,(0)}=0,
\hspace{.4cm}
{\cal V}_{1/m^2}^{\,(0)}\equiv v_0(\epsilon)
=-\frac{4-\epsilon-2\epsilon^2}{6-4\epsilon}.
\end{eqnarray}
The matching coefficients in PNRQCD receive contributions from 
hard momenta $k\sim m$ and soft momenta ${\bf k}\sim 
m \alpha_s$. Including counterterms (see below) the $d$-dimensional
coefficients assume the form ($\beta_0=11C_A/3-4T_F n_f/3$) 
\begin{eqnarray}
\label{vc1}
{\cal V}_{C}^{\,(1)}
&=&
   \bigg[\bigg(\frac{\mu^2}{\bf{q}^2} \bigg)^{\!\epsilon} -1
   \bigg]\, \frac{\beta_0}{\epsilon}\,
+\bigg(\frac{\mu^2}{\bf{q}^2} \bigg)^{\!\epsilon}\, a_{1}(\epsilon),
\\
{\cal V}_{1/m}^{\,(1)}
&=&
\bigg(\frac{{\mu }^2}{\bf{q}^2} \bigg)^{\!\epsilon }
\, b_{1}(\epsilon),
\\
{\cal V}_{1/m}^{\,(2)}
&=&
  \Bigg[\bigg(\frac{{\mu }^2}{\bf{q}^2} \bigg)^{\!2 \epsilon }-1
  \Bigg]\, \left(-\frac{8}{3\epsilon}\right)
           \bigg(2 C_F C_A + C_A^2 \bigg)\,
\nonumber
\\
&& +\, \Bigg[\bigg(\frac{\mu^2}{\bf{q}^2}\bigg)^{\!2\epsilon}
    - \bigg( \frac{{\mu }^2}{\bf{q}^2}\bigg)^{\!\epsilon}\,
  \Bigg]\, \frac{2\beta_0}{\epsilon} \, b_{1}(\epsilon)\,
+
  \bigg(\frac{\mu^2}{\bf{q}^2} \bigg)^{\! 2\epsilon }\,
  4 b_{2}(\epsilon),
\label{vb2}\\
{\cal V}_{1/m^2}^{(1)}
&=&
   \Bigg[\bigg(\frac{\mu^2}{\bf{q}^2} \bigg)^{\!\epsilon }-1
   \Bigg]\,
   \frac{1}{\epsilon}\,
   \bigg(\,\frac{7}{3}C_F-\frac{11}{6}C_A\, + \beta_0\, v_{
     0}(\epsilon)\bigg)
\nonumber
\\
&&+\,\Bigg[\left(\frac{{\mu }^2}{m^2} \right)^{\!\epsilon}-1
   \Bigg]\, \frac{1}{\epsilon}\,
   \bigg(\,\frac{C_F}{3}+\frac{C_A}{2}\,\bigg)
  +\bigg(\frac{\mu^2}{\bf{q}^2} \bigg)^{\!\epsilon}\,
        v^{(1)}_{q}(\epsilon)
  +\bigg(\frac{{\mu }^2}{m^2} \bigg)^{\!\epsilon}\,
        v^{(1)}_{m}(\epsilon),
\\
{\cal V}_{p}^{(1)}
&=&
   \Bigg[\bigg( \frac{\mu^2}{\bf{q}^2} \bigg)^{\!\epsilon }-1
   \Bigg]\, \frac{1}{\epsilon}\,
   \bigg(\frac{8}{3}C_A+\beta_0\bigg)
 +\bigg( \frac{\mu^2}{\bf{q}^2} \bigg)^{\!\epsilon}\,
        v^{(1)}_p(\epsilon).
\label{vp1}
\end{eqnarray}
The calculation of the coefficients ${\cal V}_{1/m}^{(2)}$, 
${\cal V}_{1/m^2}^{(1)}$ and ${\cal  V}_{p}^{(1)}$ 
results in IR divergences \cite{Kniehl:2002br,Brambilla:1999xj}.
In writing the above equations, we subtracted 
the expression 
\begin{eqnarray}
\delta\widetilde{V}_{c.t.}
&=&
\frac{\alpha_s C_F}{6\epsilon}
\Bigg[
C_A^{3}\frac{\alpha_s^{3}}{{\bf q}^2}
+ 4\left(C_A^{2}+2C_A C_F\right)\,
\frac{\pi\alpha_s^{2}}{m |{\bf q}|}
\nonumber \\
&& +16\left(C_F-\frac{C_A}{2}\right)\frac{\alpha_s}{m^2} +
16C_A\frac{\alpha_s}{m^2}\, \frac{\left({\bf p^2}+{\bf
p^{\prime\,2}}\right)}{2 {\bf q^2}} \Bigg],
\label{counter}
\end{eqnarray}
which corresponds to a $\overline{\rm MS}$ definition of the 
spin-projected potentials as PNRQCD matching coefficients. 
The counterterm (\ref{counter}) is added back to the 
ultrasoft contribution \cite{Beneke:2007bkp}, where it 
cancels a corresponding UV divergence. The terms proportional to   
$\beta_0/\epsilon$ in (\ref{vc1}) to (\ref{vp1}) 
are related to QCD charge renormalization. With the exception 
of $b_1(\epsilon)$, which we need to second order in the 
$\epsilon$-expansion, all other coefficients are needed at the first order. 
Their expressions are 
\begin{eqnarray}
a_{1}(\epsilon)
&=&
\bigg(C_A\,[11-8\epsilon]-4\,T_Fn_f\bigg)\,
\frac{e^{\epsilon\gamma_E}\,\Gamma(1-\epsilon)\,\Gamma(2-\epsilon)\,
\Gamma(\epsilon)\,
}{(3-2\,\epsilon)\,\Gamma(2-2\,\epsilon)}
-\frac{\beta_0}{\epsilon},
\\
b_{1}(\epsilon)
&=&
\left(\frac{C_F}{2}\,[1-2\epsilon]
      -C_A\,[1-\epsilon]
\right)\,
\frac{ e^{\epsilon\gamma_E}\,\Gamma(\frac{1}{2}-\epsilon)^{2}
       \Gamma(\frac{1}{2}+\epsilon)
     }{\pi^{\frac{3}{2}}\,\Gamma(1-2\epsilon)},
\\
b_{2}(\epsilon)
&=&
 \left[\frac{65}{18}-\frac{8 }{3}\ln{2}\right] C_AC_F
-\left[\frac{101}{36}+\frac{4 }{3}\ln{2}\right] C_A^{2}
\nonumber \\
&&
+ \left[\frac{49}{36}\,C_A -\frac{2}{9} C_F\right] T_F n_f
+ \epsilon \,b_{2}^{(\epsilon)},
\\
v^{(1)}_{q}(\epsilon)
&=&
-\frac{C_F}{3}
-\frac{11}{27} C_A
+\frac{40}{27} \,T_F n_f
+ \epsilon\, v^{(1,\epsilon)}_{q}\, ,
\\
v^{(1)}_{m}(\epsilon)
&=&
-\frac{C_F}{3}
-\frac{29}{9} C_A
+\frac{4}{15}\,T_F
+ \epsilon\, v^{(1,\epsilon)}_{m}\, ,
\\
v^{(1)}_{p}(\epsilon)
&=&
\frac{31}{9} C_A
-\frac{20}{9} \,T_F n_f
+ \epsilon\, v^{(1,\epsilon)}_p,
\end{eqnarray}
where we checked the $d$-dimensional expression for $a_1(\epsilon)$  
in~\cite{Schroder:1999sg}, $b_1(\epsilon)$ is from~\cite{Beneke:1999qg},
$b_2(\epsilon=0)$ from \cite{Kniehl:2001ju}, 
and $v_i^{(1)}(\epsilon=0)$ for $i=\{q,m,p\}$ are 
obtained from \cite{Kniehl:2002br,Sebastian:2003dipl}\footnote{
The coefficient $v_p^{(1)}$ corresponds to 
$-d_1^p$ in \cite{Kniehl:2002br}.}. 
The $O(\epsilon)$ term of 
$b_2(\epsilon)$ is unknown; the corresponding terms 
$v_i^{(1,\epsilon)}$ can be obtained from 
\cite{Sebastian:2003dipl}. However, in the present work, 
we shall use this result only for a rough estimate.

\section{Third-order  non-Coulomb correction to the wave function 
at the origin}

Since the relevant terms in the effective Lagrangian (\ref{pnrqcd}) 
do not take us out of the quark-antiquark sector of the 
Fock space, the calculation of $G(E)$ can be mapped to 
a Hamiltonian problem in single-particle quantum mechanics 
after separating the trivial center-of-mass  motion of the 
quark-antiquark pair. At leading order $G(E)$ equals the Green
function 
\begin{eqnarray}
G_0(E) &\equiv& \langle 0|\hat{G}_0|0\rangle =
\langle 0| \frac{1}{H_0 - E-i \epsilon} |0\rangle,
\end{eqnarray}
of the Schr\"odinger operator 
$H_0=\bff{\nabla^2}/m-\alpha_s C_F/r$ with $|0\rangle$ denoting a
relative position eigenstate with eigenvalue ${\bf r}=0$. The effect of the 
perturbation potentials $\delta V$ is taken into
account by substituting $H_0\to H=H_0+\delta V$, and then the 
Green function $G(E)$ is systematically expanded in 
powers of $\alpha_s$. The third-order non-Coulomb correction  
corresponds to the expression
\begin{eqnarray}
\delta_3 G &=& 2\, \langle 0|\hat{G}_0 \delta V_1 \hat{G}_0 \delta V_2
\hat{G}_0|0\rangle - \langle 0|\hat{G}_0\delta V_3 \hat{G}_0
|0\rangle 
\label{git}
\end{eqnarray}
with $\delta V_1$ the one-loop Coulomb potential contribution to $\delta V$, 
and  $\delta V_2$,  $\delta V_3$ the second and third order 
non-Coulomb potentials in momentum space given by
\begin{eqnarray}
\delta \widetilde V_2 &=& -\frac{{\bf p^4}}{4 m^3}\,(2\pi)^{d-1}\,\delta^{(d-1)}({\bf
  p}-{\bf p'}) +\delta \widetilde V_{1/m}^{(1)} +\delta \widetilde
  V_{1/m^2}^{(0)} + \delta \widetilde V_{p}^{(0)},
\\
\delta \widetilde V_3 &=&
 \delta \widetilde V_{1/m}^{\,(2)}
+\delta \widetilde V_{1/m^2}^{\,(1)} 
+\delta \widetilde V_{p}^{\,(1)},
\end{eqnarray}
where the first term in $\delta \tilde V_2$ is due to the relativistic
kinetic energy 
correction in (\ref{pnrqcd}), and $\delta \tilde V_X^{(k)}$ refers 
to the corresponding terms in (\ref{vc1}) to (\ref{vp1}) including 
the prefactors in (\ref{eq:potential}). The result of the 
potential insertions (\ref{git}) is matched to the pole
structure of $G(E)$, which allows us to extract the bound-state energy
and its wave function at the origin. 
The technical details of the calculation will be discussed elsewhere. 
Here we briefly summarize the main results. 

The expression (\ref{git}) should be considered as an
infinite sum of loop diagrams defined in dimensional regularization. 
Since the $(d-1)$-dimensional Coulomb Green function $\hat{G}_0$ is not known 
in closed form, one must separate the 
divergent subgraphs. Since these contain only a finite number of 
loops and since the divergent part is local, one calculates 
these subgraphs in $d$ dimensions and factorizes the pole 
part from the remainder of the diagram. All non-divergent 
diagrams can be computed in $d=4$ using the known three-dimensional
Coulomb Green function. For the divergent part of (\ref{git}) 
we find 
\begin{eqnarray}
\delta_3 G_{div}
&=&
\Bigg[
-\frac{1}{\epsilon^2}\,
  \bigg(\,
       \frac{7}{72}C_F^{2}
       +\frac{2}{9}C_A^{2}
       +\frac{23}{48}C_A C_F
       +\beta_0\,\bigg(\frac{C_A}{24}+\frac{C_F}{36}\bigg)
  \bigg)
\nonumber
\\
&&-  \frac{1}{\epsilon}\,
\Bigg\{
 \left(\frac{11}{24}
      -\frac{L_m}{12}
 \right)C_F^{2}
+
 \left(\frac{427}{324}
      -\frac{4\ln{2}}{3}
      -\frac{L_m}{8}
 \right)C_AC_F
-
 \left(\frac{5}{216}
      +\frac{2\ln{2}}{3}
 \right)C_A^{2}
\nonumber
\\
&&
+
 \left(\frac{C_A}{24}
      +\frac{C_F}{54}
 \right)\beta_0
-
 \left(\frac{1}{30}
      -\frac{29\,n_f}{162}
 \right)C_F T_F
+
\frac{49}{216}C_A T_F n_f 
\Bigg\}
\Bigg]\,
\frac{\alpha_s^3 C_F}{\pi}\,
\langle 0|\hat{G}_0|0\rangle
\nonumber
\\
&& - \Bigg[\frac{1}{4}C_A
      +\frac{1}{6}C_F
\Bigg]\,
\frac{\alpha_s^{2}C_F}{\epsilon}\,
\langle 0|
\hat{G}_0\delta V_1\hat{G}_0
|0\rangle
\end{eqnarray}
with $L_m=\ln(\mu/m)$. 
In the residue $Z_n$ the pole part in the last line cancels 
against a corresponding term in the contribution $2 c_v^{(2)} f_1$ 
in (\ref{z3}), while the divergent part proportional to 
$G_0(E)$ combines with similar terms in the ultrasoft contribution, 
and must cancel with $2 c_v^{(3)}$. From the finite part of 
$\delta_3 G$, we extract the non-Coulomb contribution to $f_3$, 
and obtain
\begin{eqnarray}
\frac{f_{3}^{nC}}{64\pi^2}
&=&
\Bigg[\,  \frac{7}{6} C_F^{3}
        + \frac{37}{12} C_AC_F^{2}
        + \frac{4}{3}C_A^{2}C_F
        + \beta _0
          \bigg(\,\frac{4}{3}C_F^{2}
                  + 2C_AC_F\bigg) 
\Bigg]\, L^{2}
\nonumber \\
&&\hspace*{-1cm} 
+ \, \Bigg[
 \, C_F^{3}\,
 \bigg(-\frac{3}{2}
         +\frac{14}{3n}
         -\frac{7S_1}{3}
 \bigg) 
+  C_A C_F^{2}\,
 \bigg(\, \frac{226}{27}
         +\frac{8\ln{2}}{3}
         +\frac{37}{3n}
         -\frac{5}{3n^2}
         -\frac{37S_1}{6}
 \bigg)
\nonumber \\
&& \hspace*{-0.3cm}+\,C_A^{2}C_F\,
 \bigg(\, \frac{145}{18}
         +\frac{4\ln{2}}{3}
         +\frac{16}{3n}
         -\frac{8S_1}{3}\,
 \bigg)
+ C_F^{2}\,T_F
  \bigg(\frac{2}{15}
         -\frac{59}{27}\,n_f
  \bigg)
\nonumber
\\
&& \hspace*{-0.3cm}
-\, \frac{109}{36}C_AC_FT_F\,n_f
+ \beta_0\,
  \Bigg\{C_F^{2} \,
            \bigg(
                  \frac{16}{3}
                 +\frac{10}{3n}
                 -\frac{75}{16n^2}
                 -\frac{\pi^2n}{9}
                 -\frac{4S_1}{3}
                 +\frac{2nS_2}{3}
            \bigg)
\nonumber
\\
&& \hspace*{-0.3cm}
+\, C_A C_F\,
            \bigg(
                  \frac{15}{8}
                 +\frac{5}{n}
                 -\frac{\pi^2 n}{6}
                 -2S_1
                 + nS_2
            \bigg)
   \Bigg\}\,
\Bigg] \, L
+
\Bigg[\,\frac{1}{3}C_F^{3}
       +\frac{1}{2}C_AC_F^{2}\,\Bigg]\,
L_{m} L
\nonumber
\\
&&\hspace*{-1cm} +\,
\Bigg[\frac{1}{12}C_F^{3}
       +\frac{1}{8} C_AC_F^{2}\,\Bigg] \,
L_m^{2}
+
\Bigg[
   C_F^{3}\,
   \bigg(
          \frac{1}{12}
         +\frac{2}{3n}
         -\frac{S_1}{3}
   \bigg)
+ C_AC_F^{2}\,
   \bigg(
         -\frac{5}{9}
         +\frac{1}{n}
         -\frac{S_1}{2}
  \bigg)
\nonumber
\\
&& \hspace*{-0.3cm}
+\, \frac{1}{15}C_F^{2}T_F
\Bigg]\, L_m 
+
\frac{ c_{\psi,3}^{nC}}{64\pi^2},
\end{eqnarray}
where $L\equiv\ln(n\mu/(m\,C_F\,\alpha_s))$, and the unwritten 
argument $n$ 
of the harmonic sum $S_a = S_a(n) \equiv \sum_{k=1}^{n} 1/k^a$ is the 
principal quantum number $n$. The non-logarithmic 
part\footnote{
  For some of the terms in our calculation, we obtained  
  analytic expressions which can be evaluated only 
  for any given value of $n$.
  To obtain the presented result for general $n$, we made a rather 
  general ansatz for the possible dependence on $n$ such as 
  harmonic sums and determined the 
  rational coefficients of the various structures from the analytic 
  result for $n=1,2,\ldots, 8$. The correctness of this result was
  then checked for all further $n$ up to $n=30$.} is given by
\begin{eqnarray}
\frac{c_{\psi, 3}^{nC}}{64\pi^2}
&=&
\Bigg[-\frac{137}{36}
      -\frac{49\pi^2}{432}
      -\frac{25}{6 n}
      +\frac{35}{12 n^2}
      + S_1 \,\bigg(\frac{3}{2}
                   -\frac{14}{3 n}
                   +\frac{7 S_1}{6}
              \bigg)
      -\frac{7 S_2}{6}
\Bigg] 
C_F^{ 3}
\nonumber
\\
&& \hspace{-1cm} +\,
\Bigg[\,
      \frac{7061}{486}
     -\frac{50\pi^2}{81}
     +\frac{1475}{108n}
     +\frac{\pi^2}{9n}
     -\frac{321}{32n^2}
     + \ln{2}\,
       \bigg(\frac{353}{54}
            +\frac{16}{3n}
            -\frac{16\ln{2}}{9}
       \bigg)
\nonumber
\\
&&  -\,S_1\,
      \bigg(\,\frac{226}{27}
             +\frac{8\ln{2}}{3}
             +\frac{37}{3n}
             +\frac{1}{n^2}
             -\frac{37S_1}{12}
      \bigg)
   -S_2\, 
      \bigg(\frac{37}{12}
             +\frac{2}{3n}
      \bigg)
\Bigg]
C_AC_F^{2}
\nonumber
\\
&& \hspace{-1cm}
+\,\Bigg[
       \frac{3407}{432}
      -\frac{5\pi^2}{18}
      +\frac{133}{9n}
      +\ln{2}\,
       \bigg(
             \frac{187}{108}
            +\frac{8}{3n}
            -\frac{8\ln{2}}{9}
       \bigg)
      -\frac{4S_2}{3}
\nonumber
\\
&&       - \,S_1 
       \bigg(\frac{145}{18}
+\frac{4\ln{2}}{3}
              +\frac{16}{3n}
              -\frac{4S_1}{3}
       \bigg)
\Bigg] 
C_A^{2}C_F
+
\Bigg[
       \frac{1}{15}
      +\frac{4}{15 n}
      -\frac{2S_1}{15}
\Bigg]
C_F^{2}T_F
\nonumber
\\
&& \hspace{-1cm} 
+\,\Bigg[-\frac{361}{108}
      +\frac{49\ln{2}}{108}
      -\frac{109}{18n}
      + \frac{109S_1}{36}
\Bigg] 
C_AC_FT_F\,n_f
\nonumber
\\
&& \hspace{-1cm} 
+\,\Bigg[
      -\frac{3391}{486}
      +\frac{5\pi^2}{648}
      -\frac{2\ln{2}}{27}
      -\frac{118}{27n}
      +\frac{125}{24n^2}
      + \frac{59S_1}{27}
\Bigg] 
C_F^{2}T_F\,n_f
\nonumber
\\
&& \hspace{-1cm} 
+\,\beta_0\,\Bigg[\,
\Bigg\{
    \frac{1027}{648}
   +\frac{19}{6 n}
   +\frac{25}{24n^2}
   -\frac{35\pi^2}{108}
   -\frac{11\pi^2n}{27}
   +\frac{5\pi^2}{16n}
   +\frac{4nS_3}{3}
   -\frac{2n S_{2,1}}{3}
\nonumber
\\
&& \hspace*{0.5cm}-\,S_1
   \bigg(\frac{10}{9}
         +\frac{1}{3n}
         +\frac{45}{16n^2}
         -\frac{\pi^2 n}{9}
         +\frac{2nS_2}{3}
   \bigg)
   + S_2
    \bigg(1
         +\frac{22n}{9}
         -\frac{15}{8n}
   \bigg) 
\Bigg\}\,
C_F^{ 2} 
\nonumber
\\
&&+\,\Bigg\{
    \frac{7}{24}
   -\frac{91\pi^2}{144}
   -\frac{1}{4n}
   -\frac{5\pi^2n}{24}
   - S_1 
   \bigg(\frac{3}{8}
        +\frac{1}{2n}
        -\frac{\pi^2n}{6}
        + nS_2
   \bigg)
  + S_2
   \bigg(
      \frac{3}{2}
     +\frac{5n}{4}
   \bigg)
\nonumber
\\
&& \hspace*{0.5cm} -\,n S_{2,1}
  + 2 n S_3
\Bigg\}\,
C_A C_F
\Bigg] 
+
\delta_{\epsilon},
\end{eqnarray}
where $S_{a,b}=S_{a,b}(n)\equiv\sum_{k=1}^{n} S_b(k)/k^a$ is a nested 
harmonic sum, and
\begin{eqnarray}
\delta_{\epsilon}=
\left(
       \frac{v^{(1,\epsilon)}_{m} }{8}
      +\frac{v^{(1,\epsilon)}_{q}}{12}
      +\frac{v^{(1,\epsilon)}_p}{12}\,\right) C_F^{2}\,
-\frac{C_F}{6}\,b_{2}^{(\epsilon)},
\label{epspart}
\end{eqnarray}
denotes the contribution from the unknown $O(\epsilon)$ terms of
the potentials. Note that $\delta_{\epsilon}$ is independent of 
the principal quantum number $n$.

\section{Numerical results}

We briefly discuss the numerical significance of the non-Coulomb
correction. The numbers below
should be considered as indicating the typical size of the 
correction. The correction itself is strongly scale-dependent 
(as well as scheme-dependent due to (\ref{counter}))  
and a physically relevant result is only obtained by combining 
all terms in (\ref{z3}). This cannot be done, since 
$c_v^{(3)}$ and $\delta_\epsilon$ are still unknown.
 
The numerical expressions
of the non-Coulomb wave function correction to the lowest three
states $n=1,2,3$ are given by ($\delta_3 |\psi_n(0)|^2_{nC}
=|\psi_n^{(0)}(0)|^2 (\alpha_s/(4\pi))^3 f_{3}^{\,nC}$)
\begin{eqnarray}
\delta_3 |\psi_1(0)|^2_{nC}
&=&
\frac{\left(m\alpha_sC_F\right)^3}{8\pi}
\,\frac{\alpha_s^{3}}{\pi}\,
\Big[
 \left(149.3-6.9 n_f\right) L^2
+ 0.9 L_m^{2}
+ 3.5 L \,L_m
\nonumber \\
&&
+ \left(449.8-21.9 n_f\right) L
+ 0.8 L_m
+ \left(-149.7-3.1 n_f\right)
+ \delta_{\epsilon}
\Big] ,
\label{n1num}\\
\delta_3 |\psi_2(0)|^2_{nC}
&=&
\frac{\left(m\alpha_sC_F\right)^3}{64\pi}
\,\frac{\alpha_s^{3}}{\pi}\,
\Big[
 \left(149.3-6.9 n_f\right) L^2
+ 0.9 L_m^{2}
+ 3.5 L \,L_m
\nonumber \\
&&
+ \left(211.7-13.5 n_f\right)L
- 4.4  L_m
+ \left(-217.9+2.7 n_f\right)
+ \delta_{\epsilon}
\Big] ,
\\
\delta_3 |\psi_3(0)|^2_{nC}
&=&
\frac{\left(m\alpha_sC_F\right)^3}{216\pi}
\,\frac{\alpha_s^{3}}{\pi}\,
\Big[
 \left(149.3-6.9 n_f\right) L^2
+ 0.9 L_m^{2}
+ 3.5 L \,L_m
\nonumber \\
&&
+ \left(89.7-8.8 n_f\right)L
-6.7 L_m
+\left(-218.3+5.1 n_f\right)
+\delta_{\epsilon}
\Big] ,
\end{eqnarray}
Since some of the $O(\epsilon)$ parts of the potentials, 
$\delta_\epsilon$, are not known, we perform a crude estimate. 
We derive the expressions for $v_i^{(1,\epsilon)}$, 
$i=m,q,p$ from \cite{Sebastian:2003dipl}, and estimate 
$b_2^{(\epsilon)}/b_2(\epsilon=0)$ by the corresponding 
ratio for $b_1(\epsilon)$, allowing for a factor $\pm 2$ 
variation in this ratio. In this way we find that $\delta_\epsilon$ lies 
between $-3$ and $+8$, is dominated by $b_2^{(\epsilon)}$, 
but is more than an order of magnitude smaller than  the 
other non-logarithmic terms. Below, we therefore set 
$\delta_\epsilon$ to zero. 
The potential scale is of order $m \alpha_s C_F$, while the natural choice
for the hard factorization scale is $m$. In the following 
we show the size of the non-Coulomb correction for two 
scale choices, $\mu_B = mC_F\alpha_s(\mu_B)$ and $\mu = m$.

\paragraph{\it ``Toponium'' wave function at the origin.}
For the top-antitop production cross section near threshold, 
the height of the smeared resonance is related to the wave function 
at the origin of the $n=1$ ground state. The two scales 
discussed above correspond to $\alpha^{(n_f=5)}_s(\mu_B)=0.140$
($\mu_B=32.6\,{\rm GeV}$) and $\alpha^{(n_f=5)}_s(m_t)=0.107$
($m_t=175 \,{\rm GeV}$). From (\ref{n1num}), we obtain
\begin{eqnarray}
\frac{\delta_3 |\psi_1(0)|_{nC}^2
    }{|\psi_1^{(0)}(0)|^2}
&=&
\frac{\alpha_s^3(\mu_B)}{\pi}\,
\left(-165.1+0.8\,\ln\,( \alpha_sC_F)+0.9\,\ln^2( \alpha_sC_F)\,\right)
=
-0.14,
\\
\frac{\delta_3 |\psi_1(0)|_{nC}^2
     }{|\psi_1^{(0)}(0)|^2}
&=&
\frac{\alpha_s^3(m_t)}{\pi}\,
\left(-165.1+340.3 \ln\left(\frac{1}{\alpha_s C_F}\right)+
114.7\ln^2\!\left(\frac{1}{\alpha_s C_F}\right)\right)
=0.36,\qquad
\end{eqnarray}
where $\mu=\mu_B$ ($\mu=m_t$) 
is used in the first (second) line. 
Our estimate for $\delta_\epsilon$ affects these numbers 
by less than 0.01. For comparison, the NNNLO Coulomb correction is 
$\delta_3 |\psi_1(0)|_{C}^2/|\psi_1^{(0)}(0)|^2=-0.04$ 
with $\mu=\nu=\mu_B$ (see~\cite{Beneke:2005hg}).
The result above indicates a potentially very large 
third-order contribution from the non-Coulomb potentials. 
A similar observation is made for the ultrasoft 
contribution~\cite{Beneke:2007bkp}, and since the non-logarithmic 
terms of the non-Coulomb and ultrasoft contribution have opposite 
sign, a cancellation is possible.

\paragraph{\it Bottomonium.} Here the QCD corrections to the wave function
at the origin are known to be rather large, 
and the application of the perturbative calculation
is certainly questionable for excited states~\cite{Beneke:2005hg}. 
Therefore, mainly the bottomonium ground state 
$\Upsilon(1S)$ is of  phenomenological interest. 
The bound-state scale corresponds to $\alpha^{(n_f=4)}_s(\mu_B)=0.30$ 
($\mu_B=2\,{\rm GeV}$), and the hard scale to 
$\alpha^{(n_f=4)}_s(5\,{\rm GeV})=0.21$ ($m_b=5\,$GeV).
The non-Coulomb correction to the $\Upsilon(1S)$ wave function at the
origin is
\begin{eqnarray}
\frac{\delta_3 |\psi_1(0)|_{nC}^2
    }{|\psi_1^{(0)}(0)|^2}
&=&
\frac{\alpha_s^3(\mu_B)}{\pi}\,
\left(-162.0+ 0.8\,\ln( \alpha_sC_F)+0.9\,\ln^2(\alpha_sC_F )\,\right)
=
-1.4,
\\
\frac{\delta_3 |\psi_1(0)|_{nC}^2
    }{|\psi_1^{(0)}(0)|^2}
&=&
\frac{\alpha_s^3(m_b)}{\pi}\,
\left(-162.0+362.2\ln\left(\frac{1}{\alpha_s C_F}\right)+
121.6\ln^2\!\left(\frac{1}{\alpha_s C_F}\right)\right)
=1.5.\qquad
\end{eqnarray}
Once again, the contribution of $\delta_\epsilon$ does not seem to be 
important. For comparison, the NNNLO Coulomb correction is $\delta_3
|\psi_1(0)|_{C}^2/|\psi_1^{(0)}(0)|^2=-0.47$ for $\mu=\mu_B$. In view
of the size of this third-order correction the validity of a
perturbative treatment of the leptonic $\Upsilon(1S)$ width relies 
on cancellations that might occur in the complete third-order correction.

\section{Summary}
We computed the third-order correction to the $S$-wave quarkonium wave
functions at the origin originating from insertions 
involving potentials other than the Coulomb potential. 
The UV as well as IR divergences are regulated dimensionally 
in a way that is consistent with other parts of the calculation, 
performed in the context of the diagrammatic 
threshold expansion. Together with the previous calculation 
of the Coulomb potential 
contributions~\cite{Beneke:2005hg,Penin:2005eu} and the 
ultrasoft contribution~\cite{Beneke:2007bkp} this completes 
the bound-state part of the third-order calculation. The 
remaining unknowns are related to the finite parts of 
three-loop matching coefficients 
($a_3$, $c_v^{(3)}$) and the $O(\epsilon)$ parts of 
one- and two-loop coefficients (parameterized here 
by $\delta_\epsilon$). Numerical estimates of the new 
non-Coulomb contribution suggest important 
third-order effects even for top quarks, but a definitive 
conclusion can only be drawn once the missing matching coefficients 
are computed. 

\vspace*{1em}

\noindent
\subsubsection*{Acknowledgement}
This work is supported by the DFG Sonder\-forschungsbereich/Transregio~9
``Computergest\"utzte Theoretische Teilchenphysik'' and the 
DFG Graduiertenkolleg ``Elementar\-teil\-chen\-physik an der TeV-Skala''.



\begin{thebibliography}{99}

\bibitem{Caswell:1985ui}
W.~E.~Caswell and G.~P.~Lepage,
Phys.\ Lett.\  B {\bf 167} (1986) 437.

\bibitem{Pineda:1997bj}
A.~Pineda and J.~Soto,
Nucl.\ Phys.\ Proc.\ Suppl.\  {\bf 64} (1998) 428
[hep-ph/9707481].

\bibitem{Luke:1999kz}
M.~E.~Luke, A.~V.~Manohar and I.~Z.~Rothstein,
Phys.\ Rev.\  D {\bf 61} (2000) 074025
[hep-ph/9910209].

\bibitem{Beneke:1997zp}
M.~Beneke and V.~A.~Smirnov,
Nucl.\ Phys.\  B {\bf 522} (1998) 321
[hep-ph/9711391].

\bibitem{Kniehl:1999ud}
B.~A.~Kniehl and A.~A.~Penin,
Nucl.\ Phys.\ B {\bf 563} (1999) 200
[hep-ph/9907489].

\bibitem{Kniehl:2002br}
B.~A.~Kniehl, A.~A.~Penin, V.~A.~Smirnov and M.~Steinhauser,
Nucl.\ Phys.\ B {\bf 635} (2002) 357
[hep-ph/0203166].


\bibitem{Beneke:2005hg}
M.~Beneke, Y.~Kiyo and K.~Schuller,
Nucl.\ Phys.\  B {\bf 714} (2005) 67
[hep-ph/0501289].

\bibitem{Penin:2005eu}
A.~A.~Penin, V.~A.~Smirnov and M.~Steinhauser,
Nucl.\ Phys.\  B {\bf 716} (2005) 303
[hep-ph/0501042].

\bibitem{Melnikov:1998ug}
K.~Melnikov and A.~Yelkhovsky,
Phys.\ Rev.\ D {\bf 59} (1999) 114009
[hep-ph/9805270].

\bibitem{Penin:1998kx}
A.~A.~Penin and A.~A.~Pivovarov,
Nucl.\ Phys.\ B {\bf 549} (1999) 217
[hep-ph/9807421].

\bibitem{Beneke:1999qg}
M.~Beneke, A.~Signer and V.~A.~Smirnov,
Phys.\ Lett.\ B {\bf 454} (1999) 137
[hep-ph/9903260].

\bibitem{Kniehl:1999mx}
B.~A.~Kniehl and A.~A.~Penin,
Nucl.\ Phys.\  B {\bf 577} (2000) 197
[hep-ph/9911414].

\bibitem{Manohar:2000kr}
A.~V.~Manohar and I.~W.~Stewart,
Phys.\ Rev.\  D {\bf 63} (2001) 054004
[hep-ph/0003107].

\bibitem{Kniehl:2002yv}
B.~A.~Kniehl, A.~A.~Penin, M.~Steinhauser and V.~A.~Smirnov,
Phys.\ Rev.\ Lett.\  {\bf 90} (2003) 212001
[hep-ph/0210161].

\bibitem{Hoang:2003ns}
  A.~H.~Hoang,
  Phys.\ Rev.\  D {\bf 69} (2004) 034009
  [hep-ph/0307376].

\bibitem{Beneke:2007bkp}
  M.~Beneke, Y.~Kiyo and A.~A.~Penin,
  Phys.\ Lett.\  B {\bf 653} (2007) 53, 
  arXiv:0706.2733 [hep-ph].



\bibitem{Beneke:1997jm}
M.~Beneke, A.~Signer and V.~A.~Smirnov,
Phys.\ Rev.\ Lett.\  {\bf 80} (1998) 2535
[hep-ph/9712302].

\bibitem{Czarnecki:1997vz}
A.~Czarnecki and K.~Melnikov,
Phys.\ Rev.\ Lett.\  {\bf 80} (1998) 2531
[hep-ph/9712222].

\bibitem{Marquard:2006qi}
P.~Marquard, J.~H.~Piclum, D.~Seidel and M.~Steinhauser,
Nucl.\ Phys.\  B {\bf 758} (2006) 144
[hep-ph/0607168].

\bibitem{Luke:1997ys}
M.~E.~Luke and M.~J.~Savage,
Phys.\ Rev.\  D {\bf 57} (1998) 413
[hep-ph/9707313].

\bibitem{Beneke:1998jj}
M.~Beneke, in: Proceedings of the 33rd Rencontres de Moriond: Electroweak 
Interactions and Unified Theories, Les Arcs, France, 14-21 Mar 1998 
[hep-ph/9806429].

\bibitem{Brambilla:1999xj}
  N.~Brambilla, A.~Pineda, J.~Soto and A.~Vairo,
  Phys.\ Lett.\  B {\bf 470} (1999) 215
  [hep-ph/9910238].

\bibitem{Schroder:1999sg}
Y.~Schr\"oder,
Ph.D. Thesis, Universit\"at Hamburg, 1999.

\bibitem{Kniehl:2001ju}
  B.~A.~Kniehl, A.~A.~Penin, M.~Steinhauser and V.~A.~Smirnov,
  Phys.\ Rev.\  D {\bf 65} (2002) 091503
  [hep-ph/0106135].

\bibitem{Sebastian:2003dipl}
S. W\"uster, Diplom Thesis, RWTH Aachen, 2003.

\end{thebibliography}
\end{document}